\documentclass[cits, aps]{PoS}

\usepackage{multirow}
\usepackage{chemarrow}
\usepackage{amsmath}
\newcommand{\tr}{{\rm Tr}}

\title{Charmed Baryon Spectroscopy from Lattice QCD with $N_{f}=2+1+1$ flavors}

\ShortTitle{Charmed Baryon Spectroscopy}

\author{\speaker{Ra\'ul A. Brice\~no}\\
        Department of Physics, University of Washington\\
        Box 351560, Seattle, WA 98195, USA\\
        E-mail: \email{briceno@uw.edu}}
\author{{Daniel Bolton}\\
        Department of Physics, Baylor University\\
One Bear Pl \#97316, Waco, TX 76798, USA\\
        }
\author{{Huey-Wen Lin}\\
Department of Physics, University of Washington\\
Box 351560, Seattle, WA 98195, USA\\
        }

\abstract{We present preliminary results of the calculation of the positive-parity ground state charm baryon spectrum using of $N_{f}=2+1+1$ dynamical quarks. The calculation uses a relativistic heavy-quark action for the valence charm quark, clover-Wilson fermions for the valence light quarks and HISQ sea quarks.
The spectrum is calculated with a lightest pion mass around 220~MeV and two lattice spacings ($a\sim0.12$~fm and $0.09$~fm) are used to extrapolate to continuum limit.
Our preliminary results are consistent with the currently measured baryon spectrum, except for the isospin-averaged $J=1/2$ $\Xi_{{cc}}$ which is approximately $2~\sigma$ above the SELEX observed value. We predict the yet-to-be-discovered double and triple-charm baryons $\Xi_{cc}^*$, $\Omega_{cc}$, $\Omega_{cc}^*$, $\Omega_{ccc}$ to have masses 3665(42)(29)~MeV, 3694(40)(45)~MeV, 3739(35)(21)~MeV and 4782(24)(28)~MeV, respectively.
}

\FullConference{The XXIX International Symposium on Lattice Field Theory - Lattice 2011\\
July 10-16, 2011\\
Squaw Valley, Lake Tahoe, California}

\begin{document}

\section{Introduction}

In recent years, interest in charmed baryon spectroscopy has resurfaced. This excitement has been partly triggered by the first observation of a double charm baryon (SELEX) candidate $\Xi^+_{{cc}}(3520)$ \cite{SELEX1} as well as its isospin partner $\Xi^{++}_{{cc}}$(3460) \cite{SELEX2}. The SELEX Collaboration has later confirmed their observation of $\Xi^+_{{cc}}(3520)$ \cite{SELEX_confirm}, but the BABAR \cite{BABAR1}, BELLE \cite{BELLE1}, and FOCUS \cite{FOCUS} experiments have seen no evidence for either state of the isospin doublet $(\Xi^+_{{cc}},\Xi^{++}_{{cc}})$. The SELEX evidence for this doublet implies unprecedented dynamics, the most surprising of which is the 60~{MeV} mass difference between the two states. All other previously observed isospin partners have mass differences one order of magnitude smaller. A nice discussion of the theoretical status of the double charm spectrum can be found in \cite{lin1}, one remarkable feature is the fact that previous unquenched  Lattice Quantum Chromodynamics (LQCD) calculations suggest the $\Xi_{{cc}}$ mass to be approximately 100--200~MeV higher than the one observed by the SELEX collaboration \cite{latt1, latt2}.

There remain many undiscovered doubly and triply charmed baryons states. The recently upgraded Beijing Electron Positron Collider (BEPCII) detector, Beijing Spectrometer (BES-III), the LHC, and the GSI future project, the antiProton ANnihilation at DArmstadt (PANDA) experiment, will help further disentangle the heavy-baryon spectrum and resolve puzzles like the one mentioned above. LQCD calculations serve as direct first-principles theoretical input for these experiments. With this goal in mind, we present preliminary results of the first LQCD calculation of the charm baryon spectrum using $N_{f}=2+1+1$ dynamical quarks.


\section{Lattice Formulation and Details} 

In this work we use the gauge configurations generated by the MILC Collaboration \cite{MILC}, in which the $N_{f}=2+1+1$ highly improved staggered quark (HISQ) action is used for the sea quarks. The implementation of the HISQ action has shown to further reduce lattice artifacts as compared to asqtad action\cite{MILC}.
 The gauge configurations are HYP-Smeared to remove ultraviolet noise from the gauge field. The valence light-quark (up, down and strange) propagators are generated using the clover-Wilson action.

In order to systematically remove the large $\mathcal{O}((m_{c}{a})^n)$ discretization artifacts (where $m_{c}$ is the charm-quark mass) we use the following implementation of the relativistic heavy-quark action for the charm sector \cite{RHQ1, RHQ2, RHQ3,RHQ4}:
 \begin{eqnarray}
S_{{Q}}&=&\sum_{{x}}\overline{{Q}}_x\left(m_0+\gamma_0D_0-\frac{a}{2}D_0^2+\nu\left(\gamma_iD_i-\frac{a}{2}D_i^2\right)
-\frac{a}{4}c_\text{B}\sigma_{{ij}}{F}_{{ij}}
-\frac{a}{2}c_\text{E}\sigma_{{0i}}{F}_{{0i}}\right)_{xx'}{Q}_x',
\end{eqnarray}
 \begin{center}
\begin{table}
\begin{tabular}{|c|cccccccc|}
\hline
& \hspace{.1cm}$a$[fm]\hspace{.1cm}& \hspace{.1cm}$am_\pi$\hspace{.1cm}&\hspace{.1cm}$\text{L}^3\times \text{T}$\hspace{.1cm}&\hspace{.1cm}$\text{L}m_\pi$\hspace{.1cm}
&\hspace{.1cm}$am_{c1}$\hspace{.1cm}&\hspace{.1cm}$am_{c2}$\hspace{.1cm}&\hspace{.1cm}$N_\text{cfgs}$\hspace{.1cm}&\hspace{.1cm}$N_\text{props}$\hspace{.1cm} \\ \hline \hline
\multirow{1}{*}{\hspace{.2cm}\textbf{A1}\hspace{.2cm}} &$0.11948(86)(79)$&0.18850(79)(55)&$24^3\times 64$&4.5&0.901&0.872&504&2016 \\\hline
\multirow{1}{*}{\textbf{A2}} &$0.11948(86)(79)$&0.13584(79)(59)&$32^3\times 64$&4.4&0.900& 0.853&477&1908  \\\hline
\multirow{1}{*}{\textbf{B1}} &$0.0867(13)(11)$&0.14050(40)(28)&$32^3\times 96$&4.5&0.561&0.535&391&1564 \\\hline
 \multirow{1}{*}{\textbf{B2}} &$0.0867(13)(11)$&0.09950(53)(23)&$48^3\times 96 $&4.8&0.552& 0.522&392&1568
\\\hline\hline
\end{tabular}
\caption{Details of the configurations and propagators used in this work. The lattice spacings and pion masses cited above include the statistical and fitting window errors. Listed are the two bare masses of the valence charm-quarks used, the number of configurations, and  the number of measurements performed for each ensemble. }
\label{ensembles}
\end{table}
\end{center}
where ${Q}_x$ is the heavy-quark field at the site $x$, $\gamma_{\nu}$ are the Hermitian gamma matrices that satisfy the Euclidean Clifford algebra,  $\sigma_{\mu\nu}=\frac{i}{2}[\gamma_{\mu},\gamma_{\nu}]$, $D_\mu$ is the first-order lattice derivative, and $F_{\mu\nu}$ is the Yang-Mills field strength tensor. The parameters $\{m_0, \nu, c_\text{B}, c_\text{E}\}$ must be tuned to assure $\mathcal{O}((m_{c}{a})^n)$ terms have been removed. For the coefficients $c_\text{B}$ and $c_\text{E}$ we use the tree-level tadpole-improved results \cite{latt1, clover_terms} $c_\text{B}=\frac{\nu}{u_0^3}, c_\text{E}=\frac{1+\nu}{2u_0^3}$ with the tadpole factor, $u_0$, defined as 
$u_0=\frac{1}{3}\left<\sum_{P} \tr{\left({U}_{{P}}\right)}\right>^{1/4}$.

The coefficients $m_0$ and $\nu$ are simultaneously determined nonperturbatively by requiring the ratio $\frac{m_{\overline{1S}}}{m_{\Omega}}\equiv\frac{m_{\eta_c}+3m_{J/\Psi}}{4m_{\Omega}}$ to be equal to its experimental value, 0.084, and \{$\eta_c$, $J/\Psi$\} to satisfy the correct dispersion relation, ${E}_{{H}}^2=m_{{H}}^2+ {p}^2$. 
The dispersion relation is matched 
using $\eta_c$ and $J/\Psi$ energies for the at the six lowest momenta: $\frac{2\pi}{L}(0,0,0),$ $\frac{2\pi}{L}(1,0,0),$ $\frac{2\pi}{L}(1,1,0)$, $\frac{2\pi}{L}(1,1,1)$, $\frac{2\pi}{L}(2,0,0)$, $\frac{2\pi}{L}(2,1,0)$,  and their permutations.
Two charm-quark masses are used for each ensemble.
Furthermore, in order to account for further discretization systematic errors, we have performed the calculation at two lattice spacings, $a\sim(.09~\text{fm},0.12~\text{fm}$); at each lattice spacing we use two different light-quark masses corresponding to $m_\pi\sim 220~\text{MeV}, 310~\text{MeV}$. 
This is all done in order to be able to extrapolate to the physical point $\left({m_{\pi}^2}/{m_{\Omega}^2}=0.007,{m_{\overline{1S}}}/{m_{\Omega}}=0.084, a=0~\text{fm}\right)$.
The lattice spacing for the ensembles is determined by extrapolating $am_{\Omega}$ linearly in ${m_{\pi}^2}/{m_{\Omega}^2}$ to its physical value. Our course (fine) lattice spacings are within 1.5~$\sigma$ (1~$\sigma$) from the values obtained by the MILC collaboration \cite{MILC} and within 1.7~$\sigma$ (1~$\sigma$) from the determination by the HPQCD collaboration using the $\Upsilon~2S-1S$ splitting \cite{HPQCD}. 

Details of the ensembles, including our determination of the lattice spacing are listed in TABLE \ref{ensembles}. The systematic uncertainty in the pion mass presented in TABLE \ref{ensembles} due to the placement of the fitting window is calculated as the standard deviation of all fitting windows within three timeslices of the window picked for performing the fit. Although calculations are performed at a single volume, from the $\text{L}m_{\pi}$ values listed in TABLE \ref{ensembles} it is clear that these calculations fall within the p-regime, ensuring that finite volume artifacts are exponentially suppressed.

The interpolating operators used for the positive-parity ground state charmed baryons are the following \cite{UKQCD}:
\begin{center}
\begin{tabular}{|c|c|}
\hline
J=1/2 & J=3/2  \\ \hline \hline
    $ \Lambda_{{c}}=\epsilon^{{klm}}{P}^+{Q}^{{k}}_{{c}}\left({q}^{{lT}}_{{u}}\Gamma^{{A}}{q}^{m}_{{d}}\right)$  &   \\
$\Xi_{{c}}=\epsilon^{{klm}}{P}^+{Q}^{{k}}_{{c}}\left({q}^{{lT}}_{{u}}\Gamma^{{A}}{q}^{m}_{{s}}\right)$
&
$(\Sigma_{{c}}^{*})^{{j}}=
\epsilon^{{klm}}{P}^+({P}^{3/2}_{{E}})^{{ij}}{Q}^{{k}}_{{c}}\left({q}^{{lT}}_{{u}}\Gamma^{{j}}{q}^{m}_{{u}}\right),$\\
$ (\Sigma_{{c}})^{{j}}=
\epsilon^{{klm}}{P}^+({P}^{1/2}_{{E}})^{{ij}}{Q}^{{k}}f_{{c}}\left({q}^{{lT}}_{{u}}\Gamma^{{j}}{q}^{m}_{{u}}\right)$&$
(\Xi_{{c}}^{*})^{{j}}=\frac{\epsilon^{{klm}}}{\sqrt{2}}
{P}^+({P}^{3/2}_{{E}})^{{ij}}{Q}^{{k}}_{{c}}\left({q}^{{lT}}_{{u}}\Gamma^{{j}}{q}^{m}_{{s}}
+{q}^{{lT}}_{{s}}\Gamma^{{j}}{q}^{m}_{{u}}\right)$\\
$(\Xi'_{{c}})^{{j}}=\frac{\epsilon^{{klm}}}{\sqrt{2}}
{P}^+({P}^{1/2}_{{E}})^{{ij}}{Q}^{{k}}_{{c}}\left({q}^{{lT}}_{{u}}\Gamma^{{j}}{q}^{m}_{{s}}
+{q}^{{lT}}_{{s}}\Gamma^{{j}}{q}^{m}_{{u}}\right)$&$
(\Omega_{{c}}^{*})^{{j}}=
\epsilon^{{klm}}{P}^+({P}^{3/2}_{{E}})^{{ij}}{Q}^{{k}}_{{c}}\left({q}^{{lT}}_{{s}}\Gamma^{{j}}{q}^{m}_{{s}}\right)$
\\
$(\Omega_{{c}})^{{j}}=
\epsilon^{{klm}}{P}^+({P}^{1/2}_{{E}})^{{ij}}{Q}^{{k}}_{{c}}\left({q}^{{lT}}_{{s}}\Gamma^{{j}}{q}^{m}_{{s}}\right)$&$
(\Xi_{{cc}}^{*})^{{j}}=
\epsilon^{{klm}}{P}^+({P}^{3/2}_{{E}})^{{ij}}{q}^{{k}}_{{u}}\left({Q}^{{lT}}_{{c}}\Gamma^{{j}}{Q}^{m}_{{c}}\right)$\\
$(\Xi_{{cc}})^{{j}}=
\epsilon^{{klm}}{P}^+({P}^{1/2}_{{E}})^{{ij}}{q}^{{k}}_{{u}}\left({Q}^{{lT}}_{{c}}\Gamma^{{j}}{Q}^{m}_{{c}}\right)$&$
(\Omega_{{cc}}^{*})^{{j}}=
\epsilon^{{klm}}{P}^+({P}^{3/2}_{{E}})^{{ij}}{q}^{{k}}_{{s}}\left({Q}^{{lT}}_{{c}}\Gamma^{{j}}{Q}^{m}_{{c}}\right)$\\

$(\Omega_{{cc}})^{{j}}=
\epsilon^{{klm}}{P}^+({P}^{1/2}_{{E}})^{{ij}}{q}^{{k}}_{{s}}\left({Q}^{{lT}}_{{c}}\Gamma^{{j}}{Q}^{m}_{{c}}\right)$&$(\Omega_{{ccc}})^{{j}}=
\epsilon^{{klm}}{P}^+({P}^{3/2}_{{E}})^{{ij}}{Q}^{{k}}_{{c}}\left({Q}^{{lT}}_{{c}}\Gamma^{{j}}{Q}^{m}_{{c}}\right)$\\\hline
\hline
\end{tabular}
\end{center}
where ${q}_{u,d,s}$ respectively denote the up, down and strange quark annihilation operators, ${Q}_{c}$ denotes the charm-quark operator, $\{{k,l,m}\}$ are color indices, while $\{{i,j}\}$ depict polarization indices. $\left(\Gamma^{{A}},\Gamma^{{i}}\right)$ are the antisymmetric and symmetric spin matrices $({C}\gamma_5, {C}\gamma^i)$, where C is the charge conjugation matrix. In order to have the best possible overlap with the state of interest, we have used the spin projection operators $({P}^{3/2}_{E})_{ij}= \delta_{ij}-\frac{1}{3}\gamma_{i}\gamma_{j}$ and $( {P}^{1/2}_{E})_{ij}= \delta_{ij}-({P}^{3/2}_{E})_{ij}=\frac{1}{3}\gamma_{i}\gamma_{j}$, and the positive-parity projection operator ${P}^+={(1+\gamma_4)}/{2}$.

\begin{figure}[htp]
\begin{center}
\includegraphics[totalheight=9cm]{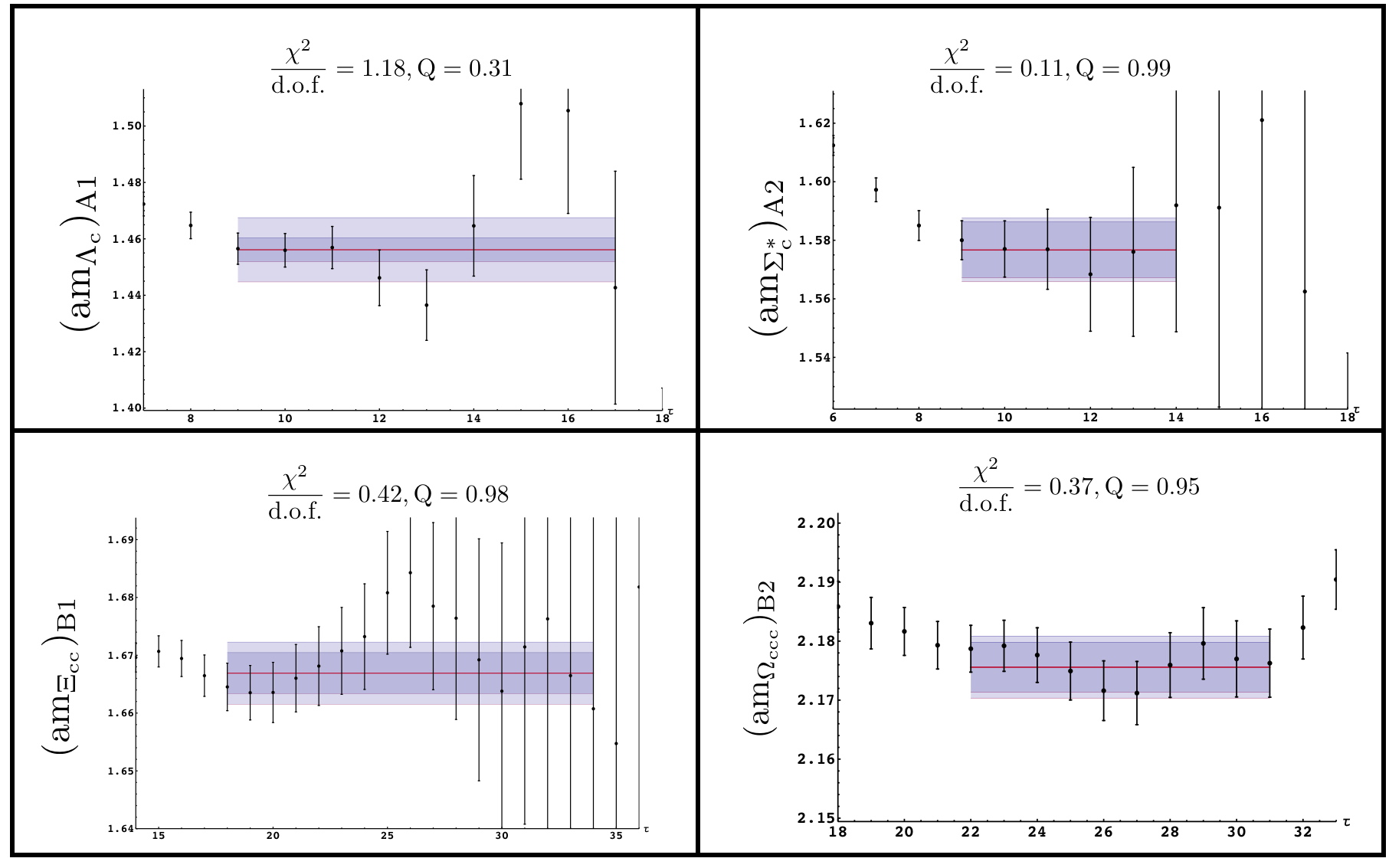}
\end{center}
\caption{Sample effective mass plots from the various ensembles. The fit includes two errorbars, the smallest of which is the statistical uncertainty, while the wider band includes the statistical and systematic uncertainty added in quadrature.  Additionally listed are $\chi^2$ per degree of freedom and the ${Q(d)}=\frac{1}{2^{d/2}\Gamma(d/2)}\int^{\infty}_{\chi^2}d\chi_0^2(\chi_0^2)^{d/2-1}e^{-\chi_0^2/2}$ factor which depends on the number of degrees of freedom $d$ and is optimally near 1.  }\label{EMPs}
\end{figure}

\begin{figure}[htp]
\begin{center}
\includegraphics[totalheight=6.80cm]{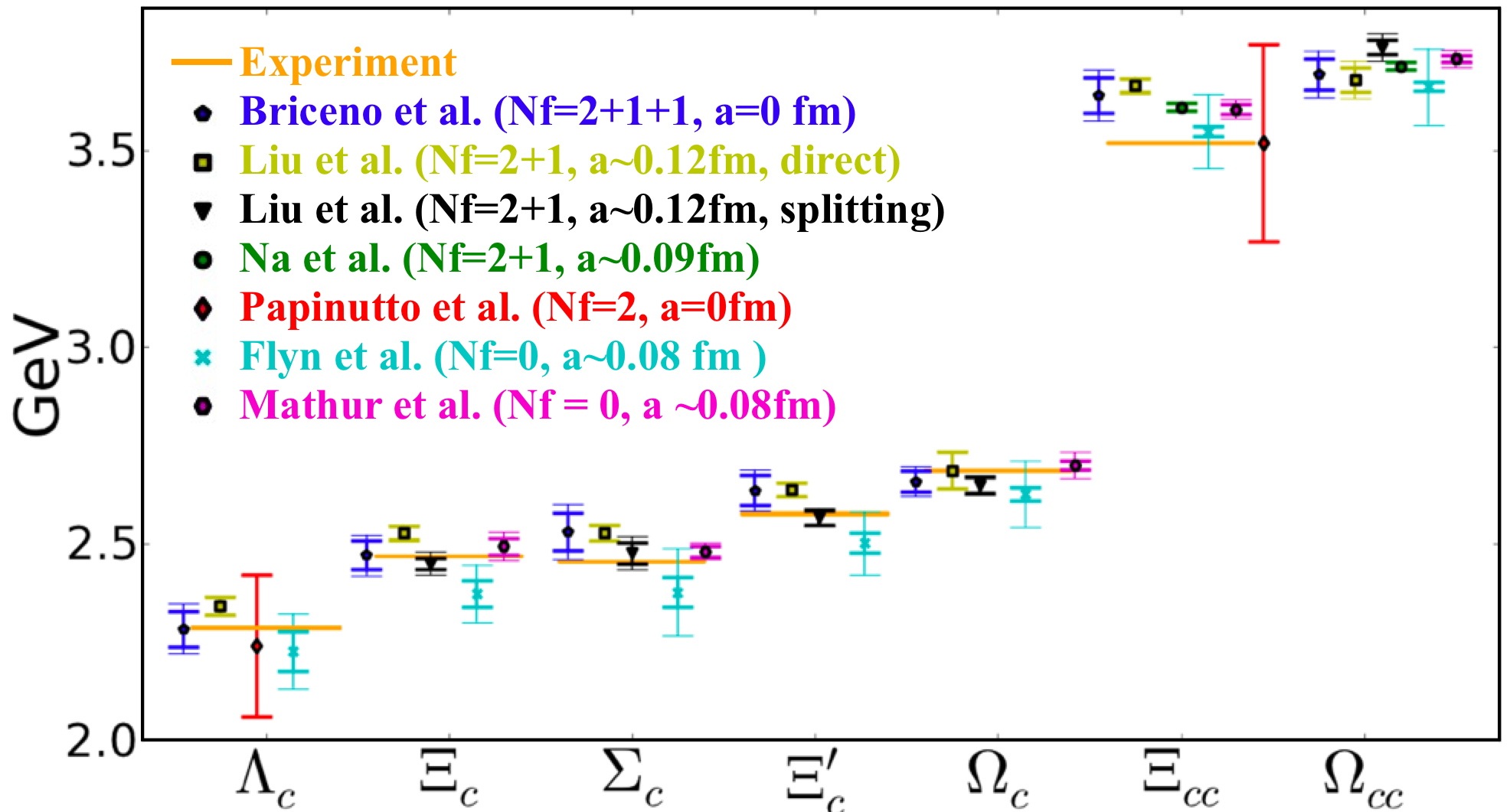}
\includegraphics[totalheight=6.85cm]{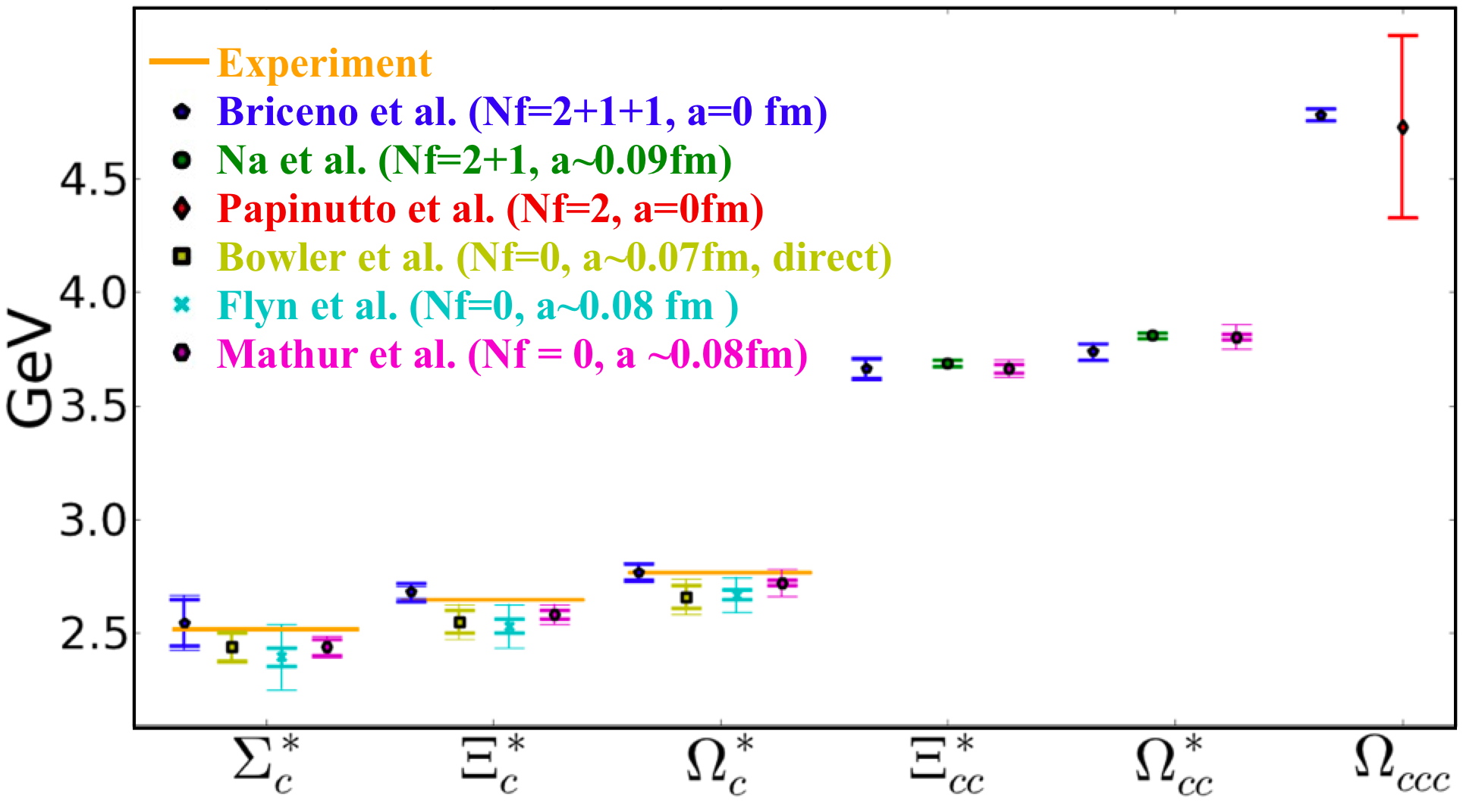}
\end{center}
\caption{Shown is a survey of previous lattice calculations \cite{latt1, latt2, latt3, latt4,latt8, latt9}, along with the results of this paper labeled as ``Briceno et al.''  The statistical uncertainty is show as a thick/inner error bar, while the statistical and systematic uncertainties added in quadrature are shown as a larger thin error bar. Our systematic uncertainties include errors originating from the fitting window and scale setting. The experimentally determined masses are listed for comparison \cite{pdg}. It is clear that our results are in agreement with the experimental values ($\leq\sigma$) with the clear exception of $\Xi_{{cc}}$, where our result is 1.9$\sigma$'s above the experimental value.}\label{Spectrum}
\end{figure}
 All two-point functions were calculated with gauge-invariant Gaussian-smeared sources and point sinks. Four sources are used in each gauge configuration.The masses were extracted from the single exponential fit to the correlation function at large Euclidean time. Double-exponential form is also used to crosscheck the ground-state mass. The uncertainties of the extracted hadron masses are evaluated using the jackknife method. Figure (\ref{EMPs}) displays examples of the effective mass plots for various correlation functions. The plots show the $\chi^2$ per degree of freedom, the Q(d) factor, along with the statistical and fitting window uncertainties for the fitting window chosen.

 \section{Results and Conclusion}


For every hadron we calculated the ratio of its mass to the $\Omega$ mass, $m_{{H}}/{m_{\Omega}}$.
The mass of each hadron is calculated at two different values of the charm-quark mass, two pion masses, and two lattice spacings. The first of these allows us to interpolate the ratio, $m_{{H}}/{m_{\Omega}}$, linearly in $am_{c}$ to the physical charm-quark mass defined by $\frac{m_{\overline{1S}}}{m_{\Omega}}= 0.084$.
We then simultaneously extrapolated the four values of the hadron masses. To perform the extrapolation we used the leading order polynomial dependance on $m_\pi$ and on the lattice spacing, $a,$ $m_{{H}}/m_{\Omega} =r_{{0}}+c_1 m_{\pi}^2/m_{\Omega}^2 +c_2a^2$.
The continuum limit hadron mass is recovered by  multiplying the physical $\Omega$ mass, 1672.45(49)~MeV.
Using this procedure, we have verified that our calculations reproduce the low-lying charmed meson spectrum. We check on the residual heavy-quark discretization effects using the mesonic spectrum after extrapolating to the physical point, and
we obtain the following results for the charmonium hyperfine splitting, D and $\text{D}_\text{s}$ masses: $\Delta_{1S}\equiv m_{J/\Psi}- m_{\eta}=111.0(5.5)(6.6)$~MeV, $m_{\text D}=1862.5(7.9)(6.7)$~MeV, $m_{{\text D}_{{s}}}$=1965(6)(11)~MeV. These are well in agreement with their corresponding experimentally observed values (116~MeV, 1869.6~MeV, 1968.5~MeV respectively). The hyperfine splitting demonstrates the necessity for continuum extrapolation when evaluating the charmed spectrum. The values of our hyperfine splitting for the course and fine ensembles are $\Delta_{1S}^{0.12}=76(12)(8)$~\text{MeV}, $\Delta_{1S}^{0.09}=95.0(4.5)(4.9)~\text{MeV}$, the finest of which underestimates the physical value by approximately 18\%. This underestimation is consistent with previous lattice calculations \cite{meson1, meson2}.

In Fig. (\ref{Spectrum}) we display our preliminary charmed baryon results, along with a survey of previous lattice calculation of the J=1/2 and J=3/2 ground state spectrum and the corresponding experimental values. 
All previous calculations of the charmed baryon spectrum have been performed with light-quark masses corresponding to $m_\pi\geq 290\text{~MeV}$, placing our calculations closest to the physical point. Additionally, from  Fig. (\ref{Spectrum}) it is clear that our calculation represents the most precise determination of the continuum spectrum of charmed baryons.
Our results are well in agreement with the experimental masses for all particles observed with the clear exception of $\Xi_{{cc}}$, where our result for the courses lattice is 1.9$\sigma$'s above the experimental value. This $\sim120$~MeV deviation from the experimentally determined mass has been observed in previous unquenched lattice calculations \cite{latt1, latt2}. As mentioned above, the experimental value of this hadron has only been observed by SELEX collaboration with no confirmation by other groups. Theoretical prediction, our results included, suggests that this experimental finding by the SELEX collaboration needs further confirmation. Lastly, we predict the yet-discovered double and triple-charm baryons masses $\Xi_{cc}^*$, $\Omega_{cc}$, $\Omega_{cc}^*$, $\Omega_{ccc}$ to be 3665(42)(29)~MeV, 3694(40)(45)~MeV, 3739(35)(21)~MeV and 4782(24)(28)~MeV, respectively.

\section*{Acknowledgments}
We thank MILC  and PNDME collaboration
for sharing their HISQ lattices and light clover propagators with us.
RB thanks Martin Savage for his feedback on the earlier draft of this proceeding.
These calculations were performed using the Chroma software
suite \cite{chroma}
on Hyak clusters at the University of Washington
eScience Institute, using hardware awarded by NSF grant
PHY-09227700. 
This work was made possible with the use of advanced computational, storage, and networking infrastructure provided by the Hyak supercomputer system, supported in part by the University of Washington eScience Institute.
The authors were supported by the DOE grant DE-FG02-97ER4014.


\begin{thebibliography}{99}

\bibliographystyle{unsrt}
\bibitem{SELEX1} M. Mattson et al. (SELEX Collaboration), Phys. Rev. Lett. {\bf 89}, 112001 (2002) [{hep-ex/0208014}].
\bibitem{SELEX2} J. Russ (SELEX Collaboration) (2002) [{hep-ex/0209075}].
\bibitem{SELEX_confirm} A. Ocherashvili et al. (SELEX Collaboration), Phys. Lett. {\bf B628}, 18 (2005) [{hep-ex/0406033}].
\bibitem{BABAR1} B. Aubert et al. (BABAR Collaboration), Phys. Rev. {\bf D74}, 011103 (2006) [{hep-ex/0605075}].
\bibitem{BELLE1} R. Chistov et al. (BELLE Collaboration), Phys. Rev. Lett. {\bf 97}, 162001 (2006) [{hep-ex/0606051}].
\bibitem{FOCUS}  S. P. Ratti (FOCUS Collaboration), Nucl. Phys. Proc.
Suppl. {\bf 115}, 33 (2003).
\bibitem{lin1} H.-W. Lin (2011), arXiv:1106.1608~[hep-lat].
\bibitem{latt1} L. Liu, H.-W. Lin, K. Orginos, and A. Walker-Loud, Phys. Rev. {\bf D81}, 094505 (2010) [{arXiv:0909.3294~[hep-lat]}].
\bibitem{latt2}  H. Na and S. Gottlieb, PoS \textbf{LATTICE2008}, 119 (2008), [arXiv:0812.1235~[hep-lat]].
\bibitem{MILC}A. Bazavov et al. (MILC collaboration), Phys. Rev. {\bf D82}, 074501 (2010) [{arXiv:1004.0342~[hep-lat]}].

\bibitem{RHQ1} A. X. El-Khadra, A. S. Kronfeld, and P. B. Mackenzie, Phys. Rev. {\bf D55}, 3933 (1997) [{hep-lat/9604004}].
\bibitem{RHQ2} S. Aoki, Y. Kuramashi and S. I. Tominaga, Prog. Theor. Phys. {\bf 109}, 383 (2003) [{hep-lat/0107009}].
\bibitem{RHQ3} N. H. Christ, M. Li, and H.-W. Lin, Phys. Rev. {\bf D76}, 074505 (2007) [{hep-lat/0608006}].
\bibitem{RHQ4} H.-W. Lin and N. Christ, Phys. Rev. {\bf D76}, 074506 (2007) [{hep-lat/0608005}].
\bibitem{clover_terms} P. Chen, Phys. Rev. {\bf D64}, 034509 (2001) [{hep-lat/0006019}].

\bibitem{HPQCD} R. J. Dowdall et al. (2011), arXiv:1110.6887~[hep-lat]. 

\bibitem{UKQCD} K.C. Bowler et al. (UKQCD Collaboration), Phys. Rev. {\bf D54}, 3619 (1996).

\bibitem{latt3} J. Flynn, F. Mescia, and A. S. B. Tariq (UKQCD Collaboration), JHEP {\bf 0307}, 066 (2003) [{hep-lat/0307025}].

\bibitem{latt4} N. Mathur, R. Lewis, and R. Woloshyn, Phys. Rev. {\bf D66}, 014502 (2002) [{hep-ph/0203253}].

\bibitem{latt8} J. Carbonell, V. Drach, and M. Papinutto (ETM Collaboration), Nucl. Phys. Proc. Suppl. {\bf 207-208}, 367 (2010).
\bibitem{latt9}M. Papinutto, J. Carbonell, V. Drach, and C. Alexandrou, PoS \textbf{LATTICE2010}, 120 (2010) [{arXiv:1012.2786~[hep-lat]}].

\bibitem{pdg} K. Nakamura et al. (Particle Data Group), J. Phys. G {\bf G37}, 075021 (2010).
\bibitem{meson1}  D. Mohler and  R. Woloshyn (2011), arXiv:1108.6093~[hep-lat].
\bibitem{meson2}  G. S. Bali, S. Collins, and C. Ehmann (2011),  arXiv:1110.2381~[hep-lat].
\bibitem{chroma} R. G. Edwards and B. Joo (SciDAC Collaboration), Nucl. Phys. Proc. {\bf 140}, 832 (2005) [{hep-lat/0409003}].

\end{thebibliography}

\end{document}